
%
\input harvmac
\overfullrule=0pt
\def\Title#1#2{\rightline{#1}\ifx\answ\bigans\nopagenumbers\pageno0\vskip1in
\else\pageno1\vskip.8in\fi \centerline{\titlefont #2}\vskip .5in}

\font\ticp=cmcsc10
%
%
\def\half{{1\over2}}

\def\apm{\alpha^{\prime}}
\def\G{\Gamma}

\def\e{\epsilon}

%
%
\lref\mk{D. Kaplan and J. Michelson, hep-th/9510053.}
\lref\guven{R. Guven, Phys. Lett. {\bf B276} (1992) 49.}
\lref\ck{C. Callan and I. Klebanov, hep-th/9511173.}
\lref\bk{C. Bachas, hep-th/9511043.}
\lref\mli{M. Li, hep-th/9510161.}
\lref\bdps{T. Banks, M. Douglas, J. Polchinski and N. Seiberg, to appear.}
\lref\rms{L. Romans, Nucl. Phys. {\bf B276 } (1986) 71.}
\lref\hrwi{P. Horava and E. Witten, hep-th/9510209.}
\lref\jpas{J. Polchinski and A. Strominger, hep-th/9510227.}
\lref\witdyn{E. Witten, hep-th/9507121.}
\lref\sixsus{P. Howe, G. Sierra and P. Townsend,
Nucl. Phys. {\bf B221 } (1983) 331.}
\lref\ov{H. Ooguri and C. Vafa, hep-th/95011164.}
\lref\dlsd{M. Duff and J. Lu, hep-th/9306052, Nucl. Phys.
{\bf B416 } (1994) 301.}
\lref\bsv{M. Bershadsky, V. Sadov and C. Vafa,
hep-th/9510225; hep-th/9511222.}
\lref\vfa{C. Vafa, hep-th/9511088.}
\lref\wsm{E. Witten, hep-th/9511030.}
\lref\mbh{A. Strominger, hep-th/9504090, Nucl. Phys. {\bf B451 } (1995) 96.}
\lref\wcs{E. Witten, Comm. Math. Phys. {\bf 121} (1989) 351;
G. Moore and N. Seiberg, Phys. Lett. {\bf B220} (1990) 422;
S. Carlip and I. Kogan, Phys. Rev. Lett. {\bf 64} (1990) 1487,
Mod. Phys. Lett. {\bf A6} (1991) 171.}
\lref\bbs{K. Becker, M. Becker and A. Strominger, hep-th/9507158.}
\lref\bst{E. Bergshoeff, E. Sezgin and P. Townsend, Phys. Lett.
{\bf 189B} (1987) 75, Ann. Phys. {\bf B185} (1988) 330.}
\lref\sen{A. Sen, hep-th/9510229, 9511026}
\lref\cvf{C. Vafa, hep-th/9511088}
\lref\jpas{J. Polchinski and A. Strominger, hep-th/9510227.}
\lref\town{C. Hull and P.Townsend,
hep-th/9410167, Nucl. Phys. {\bf B348} (1995) 109. }
\lref\threeb{M. Duff and J. Lu,
Phys. Lett. {\bf B273} (1991) 409.}
\lref\bhole{G. Horowitz and A. Strominger,
Nucl. Phys. {\bf B360} (1991) 197.}
\lref\wbran{C. Callan, J. Harvey and A. Strominger,
Nucl. Phys. {\bf B367} (1991) 60.}
\lref\hsst{J. Harvey and A. Strominger, hep-th/9504047.}
\lref\jpd{J. Polchinski, hep-th/9510017.}
\lref\pw{J. Polchinski and E. Witten, hep-th/9510169.}
\lref\jpe{J. Dai, R. Leigh and J. Polchinski, Mod. Phys.
Lett. {\bf A4} (1989) 2073.}
\lref\ew{E. Witten, hep-th/9510135.}
\lref\hrva{P. Horava, Phys. Lett. {\bf B231} (1989) 251.}
\lref\wit{E. Witten,  Nucl. Phys. {\bf443}
(1995) 85.}
%
%
\Title{\vbox{\baselineskip12pt
\hbox{
hep-th/9512059
}
}}
{\vbox{
\centerline{}
\centerline{}\centerline {Open P-Branes}}
}

\centerline{{\ticp Andrew Strominger}
}

\vskip.1in
\centerline{\sl Department of Physics}
\centerline{\sl University of California}
\centerline{\sl Santa Barbara, CA 93106-9530}

\bigskip
\centerline{\bf Abstract}
It is shown that many of the $p$-branes of
type II string theory and $d=11$ supergravity can have
boundaries on other $p$-branes. The rules for when this
can and cannot occur are derived from charge conservation.
For example it is found that
membranes in $d=11$ supergravity and IIA string theory
can have boundaries on fivebranes. The boundary dynamics are
governed by the
self-dual $d=6$ string.
A collection of $N$ parallel fivebranes contains
$\half N(N-1)$ self-dual strings which become tensionless
as the fivebranes approach
one another.

\Date{}
%

Type II string theories contain a variety
of BPS-saturated $p$-brane solitons
carrying a variety of charges $Q^i$ \refs{\wbran,\bhole,\jpd}.
All of these are extended extremal black holes \bhole. This
means that they are
extremal members of a one-parameter family of $M^i \ge Q^i$ solutions
which, for $M^i > Q^i$, have regular event horizons and
geodesically complete,
nonsingular spacelike slices with a second asymptotic region.  Furthermore
the $M^i > Q^i$ solutions decay via Hawking emission to the
BPS-saturated $M^i = Q^i$ states. Recently there has been spectacular
progress, initiated by Polchinski, in describing the dynamics of
those $p$-branes which carry RR charge by representing them as
D-branes in a type I theory\refs{\jpd,\jpe, \hrva,\ew,\pw,\bsv,\ov,
\sen, \mli, \ck, \vfa, \wsm, \bk}. In
this paper we will rederive some of these recent results from
low-energy reasoning in a manner that will generalize to all
$p$-branes and uncover new phenomena.

Viewing $p$-branes as extended holes in spacetime naturally
leads one to consider configurations in which one $p$-brane threads
through the hole at the
core of the second $p$-brane\foot{At the extremal limit,
many of the $p$-brane
solutions are singular or strongly coupled at the core, so the spacetime
metric is not a reliable guide to the geometry. One is still however led
to consider the fate of a $p$-brane which threads a large,
smooth non-extremal
$p$-brane which subsequently evaporates down to extremality.}. For example
consider a static
configuration consisting of two like-charged, parallel NS-NS ({\it i.e.}
symmetric)
fivebranes in the
IIB theory. The metric is given by
\eqn\dblmet{ds_{10}^2= \eta_{\mu\nu}dy^\mu dy^\nu +(1 +{\apm \over |x-x_1|^2}
+{\apm \over |x-x_2|^2})\delta_{jk}dx^j dx^k,}
where $\mu,\nu=0,...5$ and $j,k=6,...9$.
This has two infinite throats located at $x=x_1$ and $x=x_2$.
Next consider a RR closed string which comes out one throat and goes
in the next:
\eqn\xcl{\eqalign {X^0&=\tau, \cr X^1 &=x_1+ (x_2-x_1)\sigma .\cr}}
The existence of such a configuration may be obstructed by
charge conservation. In particular an $S^7$ which surrounds a
RR string has a non-zero integral  $Q^{RR}=\int *H^{RR}$ where $H^{RR}$ is the
RR 3-form field strength.
This would seem to prevent strings from ending, since in that
case the $S^7$ may be contracted to a point by slipping it off the
end. However in so doing one must first pass it through the fivebrane.
Using the explicit construction of \wbran\foot{A correction to the
zero mode wave function may be found in \hsst.} it may be
seen that the low-energy effective field theory on the
fivebrane worldvolume contains a coupling
\eqn\copl{\int d^6\sigma B^{RR}_{\mu\nu}{\cal F}^{\mu\nu},}
where $B^{RR}$ is the spacetime RR Kalb-Ramond field and $\cal F$ is the
worldbrane $U(1)$ gauge field strength. This leads to
the equation of motion
\eqn\eom{d*H^{RR}= Q^{RR}\delta^8+*{\cal F}\wedge \delta^4,}
where $\delta^4$ ($\delta^8$) is a transverse 4-form (8-form) delta
function on the fivebrane (RR string) and $*\cal F$ denotes the
Hodge dual within
the worldvolume. The total integral of $d*H$ over any $S^8$ must vanish.
Consider an $S^8$ which intersects the string at only one point.
Such an $S^8$
point must intersect the fivebrane in an $S^4$.  Integrating \eom\
over the $S^8$ we find
\eqn\sein{0=Q^{RR}+\int_{S^4} *{\cal F}.}
We conclude that $H^{RR}$ charge conservation can be maintained if an
electric flux associated to the fivebrane $U(1)$ charge
emanates from the point at which the string enters the fivebrane. In
other words
the end of the string looks like a charged particle on the worldbrane.

As most easily seen from the Green-Schwarz form of the string action,
the stretched string preserves those supersymmetries
generated by spinors $\epsilon$
obeying
\eqn\wss{\G_{MN}\partial_+X^M\partial_-X^N \e = \e.}
\xcl\  and \dblmet\ together preserves one quarter of the supersymmetries
so this configuration is BPS-saturated to leading order.
At next order one must include the back reaction of the string on the
spacetime geometry and fields. Since there is no obstruction from
charge conservation
we presume that a fully supersymmetric configuration describing a
RR string stretched between two NS-NS fivebranes exists and corresponds to
a BPS state.

There is no coupling of the form \copl\ involving the NS-NS $B$
field. Charge conservation therefore prohibits fundamental IIB
strings from ending at NS-NS
fivebranes. However $SL(2,Z)$ interchanges
NS-NS and RR onebranes and fivebranes.  Hence
$SL(2,Z)$ invariance implies that a fundamental IIB string can end at
a RR fivebrane. The latter (like all the RR solitons)
can be realized as a D-brane. So this is not
a surprise: we
have reproduced results
of \refs{\jpe,\jpd}.

Next let us consider what happens as the two fivebranes
approach one another.
The mass of the stretched string is given
by a BPS bound and is a function on
the two-fivebrane moduli space. It
decreases with the string length. When the fivebranes become coincident, the
stretched string has zero length and
becomes a massless state carrying the $U(1)$ charges of
both fivebranes. The
result is therefore an $N=4$ U(2) gauge theory on the
fivebrane \refs{\jpe,\ew}.  Note that the dual relation to open
string theory is not required for this conclusion.
{}From this perspective the source of massless
gauge bosons is similar to that in \refs{\town, \wit, \mbh}:
they arise from a degenerating one-cycle which threads two horizons.

A similar story applies to the RR threebrane. Reduction of the formulae in
\refs{\bhole,\threeb} leads to  spacetime-worldbrane
couplings of the form \copl\ for  {\it both} the NS-NS and RR $B$ fields.
This is required by
$SL(2,Z)$ invariance because the threebrane acts as a source
for the self-dual
5-form and hence is itself self-dual. In \jpe\ it was shown that
fundamental
strings can end on
D-branes but here we see that D-strings may in some cases also end
on D-branes.
This dovetails nicely with S-duality of the $N=4$, $d=4$ gauge theory which
lives on the threebrane: The ends of fundamental strings are electrically
charged particles
while the ends of  D-strings are magnetically charged particles.

There may seem to be a puzzle for example for the RR 0-brane. Clearly
charge conservation will prevent (except when there is a
RR background \jpas ) a fundamental string from ending at
a 0-brane. This may seem to conflict with the picture in
\jpd\ which involves an $SU(N)$ gauge theory for $N$
0-branes coming from
strings ending  at the 0-branes. However there is not really a conflict
because our reasoning applies only to BPS states, and charge confinement in
0+1 $SU(N)$ gauge theories indeed eliminates the charged BPS states.

So far we have reproduced from a different
perspective results previously  obtained either directly
in \refs{\jpe,\jpd, \ew}, as
well as  some $SL(2,Z)$ duals of those results. Our point of
view gives the leading low-energy dynamics, but probably cannot easily
reproduce the detailed prescription given in
\refs{\jpe,\jpd} for computing {\it e.g.} finite momentum string-D-brane
scattering as in \ck .
However in considering higher $p$-branes this low-energy
perspective will lead us to new phenomena.

As a further example we consider a membrane stretched between two
fivebranes of eleven-dimensional supergravity\foot{Preliminary
observations on open membranes are made in \bst.}. (Of course reduction
of this leads to examples in the IIA theory.) Unlike it's IIB partner,
the $d=11$ (and IIA) fivebrane has chiral dynamics governed by the
$d=6$ tensor multiplet \sixsus\ containing 5 scalars and a self-dual
antisymmetric tensor field strength $\cal A$ \wbran. The membrane
worldvolume condition
for unbroken supersymmetry is \refs{\bst, \bbs}
\eqn\wmss{\G_{MNP}\e^{\alpha \beta \gamma}\partial_\alpha X^M
\partial_\beta
X^N \partial_\gamma X^P\e = \e.}
Again it is easily seen that a membrane stretched between two fivebranes
preserves one quarter of the supersymmetries at leading order.
For appropriate brane orientations the unbroken supersymmetries are
generated by spinors obeying the two chirality conditions
$\G^{016}\e=\G^{012345}\e=\e$. The
membrane can be surrounded by an $S^7$ for which there is a
nonzero value of the charge $Q^M=\int_{S^7} *F$, where $F$ here is the
spacetime 4-form
field strength. In the presence of a membrane and a fivebrane the
equation of motion for $F$, as follows from
formulae in \refs{\wbran,\guven, \mk}, is
\eqn\eomf{d*F= Q^M\delta^8+{\cal A} \wedge \delta^5.}
We see that the boundary of the membrane -  which is a string lying in
the fivebrane - must carry self-dual antisymmetric tensor charge
$\int_{S^3}{\cal A}=-Q^M$.
This string is the self-dual string of Duff and Lu \dlsd .

Further insight into this construction can be gained by considering
S- and T- duality. Polchinski \refs{\jpe,\jpd} has shown that the
worldbrane dynamics of the IIB RR fivebrane are described by
open fundamental Dirichlet strings. $SL(2,Z)$ invariance then implies that
worldbrane dynamics of the IIB NS-NS fivebrane are described by
open RR strings (although this description is weakly coupled only at large
$g_s$). Now periodically identify and
T-dualize along one direction of the fivebrane.  This gives a IIA
theory \jpe. The NS-NS ({\it i.e.} symmetric) fivebrane
solution is represented by a conformal field theory involving only the
transverse coordinates, and hence is unaffected by longitudinal T-duality
(This is in contrast to RR $p$-branes, which lose (gain) a dimension
under longitudinal (transverse) T-duality.).
However the zero modes which propagate parallel to the
fivebrane are affected, and the $N=4$ $U(1)$ vector multiplet is transformed
into
an $N=4$ antisymmetric tensor multiplet. At the same time the open strings
which govern the IIB fivebrane dynamics are T-transformed into
open membranes which govern the IIA fivebrane dynamics.

Next we consider the
dynamics of $N$ parallel $d=11$ or IIA fivebranes.
When the fivebranes are separated the low energy dynamics is governed  by
a globally supersymmetric $(0,2)$ $d=6$ theory with $N$
tensor multiplets. The moduli space of the $5N$ scalars is uniquely
determined to be locally the symmetric space $T(5,N)\equiv
SO(5,N)/(SO(5)\times SO(N))$. Since this is a
chiral theory it is not possible
for extra massless fields to appear when the fivebrane positions
coincide.  However tensionless strings can and do arise,
because the tension of
a BPS string which arises as the boundary of a membrane stretched between
two fivebranes vanishes when the fivebrane coincides. These
strings transform in the adjoint of the global $SO(N)$ which acts on
the $N$ tensor multiplets. Upon $S^1$
compactification along the fivebranes, winding states of the
tensionless strings lead to the
appearance of extra massless gauge bosons which - together with
the reduced tensor multiplets which dualize to vector multiplets - fill out
a $U(N)$ gauge theory \witdyn, as predicted by T-duality.

Aspects of the preceding are quite similar to Witten's discussion
\witdyn\ of K3 compactification of IIB string theory,
whose moduli space is locally $T(5,21)$ and which contains
(5) 5+16 (anti) self-dual antisymmetric tensor fields.
In this case tensionless strings arise from threebranes wrapping
degenerating 2-cycles.  Indeed there is a dual IIA description of
this IIB compactification, in the spirit of \refs{\bdps,\bsv,\ov},
as 16 toroidally compactified symmetric fivebranes and
NS-NS orientifolds, where  the extra
5+5 antisymmetric tensor fields arise from the supergravity
multiplet \rms .  In \bdps\ it was shown that IIA on K3 is equivalent to
IIB on a D-manifold with 16 RR orientifolds and 16 RR fivebranes.
IIB S-duality converts NS-NS to RR fields, so this is S-equivalent to
a IIB configuration with 16 NS-NS orientifolds and 16 NS-NS fivebranes.
Next T-dualize this last  representation of IIA on K3
(yielding IIB on K3) along one of the noncompact directions.
This will not affect the 4-geometry which involves only NS-NS fields.
Hence IIB on K3 is equivalent to IIA on a ``$p$-manifold'' with
16 NS-NS orientifolds and 16 symmetric fivebranes. This provides
the concrete connection to \witdyn.

As pointed out in \witdyn\ the fact that
self-dual strings (or open membranes\foot{The relation
in the infrared between the self-dual
open membranes and self-dual strings may involve
Chern Simons theory as in \wcs.} )
become light as the fivebranes approach
one another suggests that supergravity might be decoupled and the
dynamics of self-dual strings and symmetric fivebranes
consistently studied in isolation from the rest of string theory.
This is also suggested by superconformal invariance of the
tensor multiplet in $d=6$ \sixsus. The relation by compactification
to superconformal
$d=4$ $N=4$ Yang-Mills makes this a particularly fascinating problem.

Further examples of $p$-branes with boundaries can be found.
It may be directly checked in the IIB theory
that charge conservation allows a threebrane to
end on a membrane in the RR fivebrane. The membrane carries magnetic charge
with respect to the fivebrane $U(1)$ gauge field. In general every
RR $p$-brane has a $U(1)$ gauge field. Electric charges are always
carried by zerobranes and
arise from fundamental strings which terminate at the $p$-brane.
Magnetic charges are carried by a $(p-3)$-brane, which can arise
as the boundary of a $(p-2)$-brane. It is difficult to check
charge conservation directly for $p>5$ because the zero mode
wave functions have not been worked out. However T-duality
along a dimension transverse to an configuration of RR $p$-branes
increases $p$, so we presume it is always possible
(in IIA or IIB) for a RR $(p-2)$-brane to end at a RR $p$-brane.
All of these new multi-$p$-brane configurations can be used to
construct $p$-manifold  generalizations of the
D-manifolds introduced in \bsv , and may arise in the
process of dualization.

In conclusion string theory contains a rich variety of extended
objects which interact in an intricate and beautiful fashion. Higher
$p$-branes provide endpoints for branes of lower $p$, which
latter in turn govern the dynamics of the former.

\bigskip
\centerline{\bf Acknowledgements}\nobreak

We thank J. Polchinski for useful discussions
and for explaining the results of \bdps\ prior to publication.
This work was supported in part by DOE Grant No. DOE-91ER40618.

\listrefs

\end